\documentclass[prd,aps,a4,twocolumn,superscriptaddress,notitlepage,preprintnumbers,nofootinbib]{revtex4-1}
\usepackage{pslatex}
\usepackage[pdftex]{graphicx}
\usepackage{psfrag}
\usepackage{epsfig}
\usepackage[usenames,dvipsnames]{color}
\usepackage{cancel}
\usepackage{slashed}
\usepackage{amssymb}
\usepackage{lipsum}
\usepackage{amsmath}
\usepackage{enumerate}
\usepackage{multirow}
\usepackage{natbib}
\usepackage[colorlinks=true,citecolor=darkred,urlcolor=darkred,pdfborder={0 0 0}]{hyperref}
\usepackage[normalem]{ulem}

\definecolor{darkred}{rgb}{0.6,0,0}
\definecolor{dbrown}{rgb}{0.4,0.26,0.13}
\definecolor{linkcolor}{rgb}{0,0,0.5}

\definecolor{vdrgreen}{rgb}{0.0, 0.7, 0.0}

\begin{document}
\title{Consequences of the Dresden-II reactor data\\ for
  the weak mixing angle and new physics}%
\author{D. Aristizabal Sierra}%
\email{daristizabal@ulg.ac.be}%
\affiliation{Universidad T\'ecnica
  Federico Santa Mar\'{i}a - Departamento de F\'{i}sica\\
  Casilla 110-V, Avda. Espa\~na 1680, Valpara\'{i}so, Chile}%
\author{V. De Romeri}%
\email{deromeri@ific.uv.es}%
\affiliation {Instituto de F\'{i}sica Corpuscular, CSIC-Universitat de Val\`{e}ncia, 46980 Paterna, Spain}%
\author{D. K. Papoulias}%
\email{d.papoulias@uoi.gr}%
\affiliation{Department of Physics, University of Ioannina GR-45110
  Ioannina, Greece}%
\begin{abstract}
  The Dresden-II reactor experiment has recently reported a suggestive
  evidence for the observation of coherent elastic neutrino-nucleus
  scattering, using a germanium detector.  Given the low recoil energy
  threshold, these data are particularly interesting for a low-energy
  determination of the weak mixing angle and for the study of new
  physics leading to spectral distortions at low momentum transfer.
  Using two hypotheses for the quenching factor, we study the impact
  of the data on: (i) The weak mixing angle at a renormalization scale
  of $\sim 10\,\text{MeV}$, (ii) neutrino generalized interactions
  with light mediators, (iii) the sterile neutrino dipole portal. The
  results for the weak mixing angle show a strong dependence on the
  quenching factor choice.  Although still with large uncertainties,
  the Dresden-II data provide for the first time a determination of
  $\sin^2\theta_W$ at such scale using coherent elastic
  neutrino-nucleus scattering data.  Tight upper limits are placed on
  the light vector, scalar and tensor mediator scenarios.  Kinematic
  constraints implied by the reactor anti-neutrino flux and the
  ionization energy threshold allow the sterile neutrino dipole portal
  to produce up-scattering events with sterile neutrino masses up to
  $\sim 8\,$MeV. In this context, we find that limits are also
  sensitive to the quenching factor choice, but in both cases
  competitive with those derived from XENON1T data and more stringent
  that those derived with COHERENT data, in the same sterile neutrino
  mass range.
\end{abstract}
\maketitle
% -------------------
% Sec: Introdcution
% -------------------
\section{Introduction}
\label{sec:intro}
Since first observed by the COHERENT collaboration in 2017
\cite{Akimov:2017ade} with a CsI detector, and subsequently in 2020
with a liquid argon (LAr) detector \cite{COHERENT:2020iec}, coherent elastic
neutrino-nucleus scattering (CE$\nu$NS) has been recognized as a
powerful tool for Standard Model (SM) measurements and beyond-the-SM (BSM)
searches. Examples of the physics cases that can be
studied range from the determination of the mean-square radii of
neutron distributions and low-energy measurements of the weak mixing
angle
\cite{Cadeddu:2017etk,Cadeddu:2020lky,Papoulias:2017qdn,Papoulias:2019txv,Miranda:2020tif,AristizabalSierra:2019zmy},
up to searches for new interactions in the neutrino sector covering a
whole spectrum of possible mediators (see
e.g. \cite{Farzan:2018gtr,Shoemaker:2017lzs,Dutta:2019eml,Farzan:2015hkd,Liao:2017uzy,Coloma:2017ncl,Coloma:2019mbs,Flores:2020lji,AristizabalSierra:2019ufd,AristizabalSierra:2019ykk,AristizabalSierra:2018eqm,Brdar:2018qqj,Lindner:2016wff,AristizabalSierra:2020zod,AristizabalSierra:2021fuc,Hurtado:2020vlj,Denton:2018xmq,Giunti:2019xpr,Chang:2020jwl,Miranda:2021kre,Flores:2021kzl}). Interestingly,
the same experimental infrastructures used for CE$\nu$NS measurements,
provide as well environments suitable for searches of new degrees of
freedom involving light dark matter
(LDM)\cite{Dutta:2019nbn,Dutta:2020vop,COHERENT:2019kwz,COHERENT:2021pvd}
and axion-like particles (ALPs)
\cite{Dent:2019ueq,AristizabalSierra:2020rom}.

Motivated by this wide range of possibilities, plans for further
CE$\nu$NS measurements are underway. They involve experiments using
reactor neutrinos (e.g. CONUS
\cite{Hakenmuller:2019ecb,Bonet:2020awv,CONUS:2021dwh}, CONNIE
\cite{Aguilar-Arevalo:2019jlr}, MINER \cite{Agnolet:2016zir}, RED-100
\cite{Akimov:2017hee}, $\nu$-cleus \cite{Strauss:2017cuu}, TEXONO
\cite{Wong:2015kgl}, vIOLETA \cite{Fernandez-Moroni:2020yyl}, SBC
\cite{SBC:2021yal} and the Dresden-II reactor experiment
\cite{Colaresi:2021kus}), measurements at COHERENT with germanium and
NaI detectors \cite{Barbeau:2021exu}, the Coherent CAPTAIN-Mills (CCM) experiment~\cite{CCM:2021leg} as well as at the European
Spallation Source (ESS) \cite{Baxter:2019mcx}. Plans to extended
measurements/searches with decay-in-flight neutrino beams such as NuMI
\cite{MINERvA:2016iqn} or LBNF \cite{DUNE:2016evb} using gaseous
targets with the directional $\nu$BDX-DRIFT are as well expected 
\cite{AristizabalSierra:2021uob,Abdullah:2020iiv}. Measurements of
CE$\nu$NS in multi-ton dark matter (DM) detectors and at RES-NOVA, using
archaeological lead, are part of the facilities in which CE$\nu$NS
will be looked for
\cite{Aprile:2015uzo,Aalbers:2016jon,Malling:2011va,RES-NOVA:2021gqp,Strigari:2009bq,AristizabalSierra:2017joc,Gonzalez-Garcia:2018dep,Dutta:2017nht,Schwemberger:2022fjl,Suliga:2020jfa}. Overall,
an international program covering the different energy windows where
CE$\nu$NS can be observed is well established.

These energy windows offer features that make them particularly
suitable for certain types of new physics searches. Pulsed
decay-at-rest (DAR) neutrino beams (such as those at the spallation neutron source and
the ESS) provide energy and timing spectra, thus making them particularly
useful in searches for flavor-dependent new physics. Decay-in-flight
(DIF) neutrino beams---instead---are rather suited for testing nuclear
physics hypotheses, due to their higher energy. Finally, given the
extremely low-energy thresholds of reactor experiments, sensitivity to
physics producing spectral distortions at low momentum transfer becomes
a main target. Arguably, the prototypical scenario in that case
corresponds to neutrino magnetic moments and transitions, for which the
differential cross section exhibits a Coloumb divergence
\cite{Vogel:1989iv}. Scenarios with light mediators, although not
leading to such pronounced spectral features, can also be tested with
reactor data.

In this regard the recent suggestive observation of CE$\nu$NS by the
Dresden-II reactor experiment \cite{Colaresi:2022obx} offers an
opportunity to systematically test the presence of such new light
mediators. The Dresden-II reactor experiment consists of a 2.924 kg
p-type point contact germanium detector (NCC-1701) operating at 0.2
$\text{keV}_\text{ee}$ and located at $\sim 10\,\text{m}$ from the
2.96 GW Dresden-II nuclear reactor. The data released follow from a
96.4 days exposure with 25 days of reactor operation outages in which
no visible CE$\nu$NS signal was observed. Analyses relying on these
data and investigating the implications of a modified Lindhard
quenching factor (QF) as well as limits on light vector mediators have
been already presented in Ref. \cite{Liao:2022hno}. These data have
been used also to place limits on a variety of new physics scenarios
including neutrino non-standard interactions (NSI), light vector and
scalar mediators and neutrino magnetic moments in
Ref. \cite{Coloma:2022avw}.

In this paper we extend upon these analyses and consider the impact of
the Dresden-II reactor data on: (i) Low-energy measurements of the weak mixing angle at
a $\mu\simeq 10\,$MeV renormalization scale, (ii) neutrino generalized
interactions (NGI) with light mediators, of which light vector and
scalar mediators are a subset, (iii) neutrino magnetic transition
couplings leading to up-scattering events (the so-called sterile
neutrino dipole portal \cite{McKeen:2010rx,Brdar:2020quo},
$\bar\nu_e + N \to F_4 + N$ with $F_4$ a heavy sterile
neutrino).\\

The remainder of this paper is organized as follows. In
Sec. \ref{sec:ngi-lm} we briefly present the physics scenarios treated in our
statistical analysis, including a short discussion on how
the weak mixing angle can affect the event rate. In
Sec. \ref{sec:parameter_space_analysis} we discuss differential event
rates, total event rates and the details of the statistical analysis
we have adopted along with our results. Finally, in
Sec. \ref{sec:conclusions} we present our summary and conclusions.
% -----------
% Section-I
% -----------
\section{CE$\nu$NS differential cross section, weak mixing angle and
  new physics scenarios}
\label{sec:ngi-lm}
In the SM the CE$\nu$NS differential cross section follows from a
t-channel neutral current process and reads
\cite{Freedman:1973yd,Freedman:1977xn}
\begin{equation}
  \label{eq:xsec_CEvNS}
  \left . \frac{d\sigma}{dE_r} \right|_\text{SM} = \frac{G_F\,m_N}{2\pi}Q_W^2 F^2(q^2)
\left(
  2 - \frac{m_N E_r}{E_\nu^2}
\right)\ ,
\end{equation}
where $G_F$ refers to the Fermi constant, $m_N$ to the nuclear target
mass, $E_r$ to recoil energy, $E_\nu$ to the incoming neutrino energy
and $Q_W$ to the weak charge coupling, that accounts for the
$Z^0$-nucleus interaction in the zero momentum transfer limit. Since
the scatterer has an internal structure, this coupling is weighted by
the nuclear weak form factor $F^2(q^2)$\footnote{For DAR and DIF
  neutrino beams the form factor plays an important role. For reactor
  neutrinos, instead, the energy regime is such that to a large degree
  $F^2(q^2)\to 1$.}. Hence, the ``effective'' coupling
$Q_W\times F(q^2)$ encapsulates the expected behavior: As the momentum transfer $q$
increases, the weak charge diminishes and so does the strength of the
interaction. Neglecting higher-order momentum transfer terms that
arise from the nucleon form factors, one explicitly has
\begin{equation}
  \label{eq:weak-cahrge}
  Q_W=Z\,g_{V,\text{SM}}^p + (A-Z)\,g_{V,\text{SM}}^n\ . 
\end{equation}
Here the proton and neutron vector couplings are dictated by the
fundamental $Z^0-\mathrm{q}$ ($\mathrm{q}=u,d$) couplings, given by
$g_{V,\text{SM}}^p=1/2 - 2\sin^2\theta_W$ and
$g_{V,\text{SM}}^n=-1/2$.  For the value of the weak mixing angle at
$\mu=m_{Z^0}$,
$\sin^2\theta_W|_{\overline{\text{MS}}}(m_{Z^0})=0.23122\pm 0.00003$
\cite{Tanabashi:2018oca}, one can easily check that the neutron
coupling exceeds the proton coupling by about a factor 10, resulting
in the $N^2=(A-Z)^2$ dependence predicted in the SM for the CE$\nu$NS
cross section. However, a fair amount of events allows for
sensitivities to $\sin^2\theta_W$. The SM
predicted value at $q=0$ (obtained by RGE extrapolation in the minimal
subtraction ($\overline{\text{MS}}$) renormalization scheme) is
\begin{equation}
  \label{eq:weak-mixing-angle-q-Eq0}
  \sin^2\theta_W(q=0) = \kappa(q=0)|_{\overline{\text{MS}}}
  \sin^2\theta_W|_{\overline{\text{MS}}}(m_{Z^0})\ ,
\end{equation}
with $\kappa(q=0)|_{\overline{\text{MS}}}=1.03232\pm 0.00029$
\cite{Kumar:2013yoa}. Variations around this value lead to
fluctuations of the predicted cross section and of the event rate
(see Sec. \ref{sec:parameter_space_analysis}). Although statistical
analyses of the weak mixing angle have been performed in the light of COHERENT data
\cite{Papoulias:2017qdn,Miranda:2020tif} and are expected to follow also
 from the electron channel at e.g. DUNE \cite{deGouvea:2019wav},
the interesting aspect of an analysis using reactor data has to
do with the different energy scale of such an indirect
measurement (compared with COHERENT or DUNE) and potentially with the
amount of data.
% --------------
% Section
% --------------
\subsection{Renormalizable NGI}
\label{sec:renormalizable-NGI}
Effective NGI\footnote{In contrast to the standard effective
  interaction jargon, here the typical energy scale has to be just
  above the MeV scale.  For reactor experiments this means
  $\Lambda>q\simeq 19\,$MeV.} were first considered by T. D. Lee and
Cheng-Ning Yang in Ref. \cite{Lee:1956qn}.  They have been as well
considered in the context of neutrino propagation in matter in
Ref. \cite{Bergmann:1999rz}.  More recently they have been considered
in the context of CE$\nu$NS analyses in Ref. \cite{Lindner:2016wff}
and within COHERENT CsI measurements in
Ref. \cite{AristizabalSierra:2018eqm}\footnote{In this Reference the
  acronym NGI, and thus the name ``neutrino generalized interactions''
  rather than generalized neutrino interactions, was introduced as to
  mimic the acronym NSI for ``neutrino nonstandard
  interactions''.}. Although the Dresden-II reactor data
can be used to analyze effective NGI, given its rather low recoil
energy threshold one could expect beforehand that better sensitivities
to NGI induced by light mediators are achievable. Note that an
analysis of this scenario in the context of multi-ton DM detectors
has been presented recently in Ref. \cite{Majumdar:2021vdw}.

Focusing on this case, the most general Lagrangian can be written
schematically as follows
\begin{equation}
  \label{eq:NGI_Lag}
  \mathcal{L}_{\nu-\mathrm{q}} = \sum_{\substack{X=S,P \\V,A,T}}
  \left[
    \overline{\nu}\,\,f_X\Gamma_X\,\nu\,X
    +
    \sum_{\mathrm{q}=u,d}
    \overline{\mathrm{q}}\,\Gamma_X\,\left(g_X^\mathrm{q} + i \gamma_5 h_X^\mathrm{q}\right)\,\mathrm{q}\,X
    \right]\ ,
\end{equation}
where
$\Gamma_X=\{\mathbb{I},i\gamma_5,\gamma_\mu,\gamma_\mu\gamma_5,\sigma_{\mu\nu}\}$
with $\sigma_{\mu\nu}=i[\gamma_\mu,\gamma_\nu]/2$, the parameters in
the quark and neutrino currents ($f_X, g^\mathrm{q}_X$ and
$h^\mathrm{q}_X$) are taken to be real and the interactions to be lepton
flavor universal. Here, $X$ refers to the field responsible for the
interaction. Integrating $X$ out leads to an effective Lagrangian that
contains, among other terms, NSI as a subset. In the absence of a
robust deviation from the SM CE$\nu$NS prediction, there is no a
priori reason for any of these interactions to be preferred over the
others. However, those involving nuclear spin (spin-dependent
interactions) are expected to produce lower event rates, in particular
in heavy nuclei \cite{Freedman:1977xn}. Dropping those couplings and
moving from quark to nuclear operators the resulting Lagrangian reads
\begin{widetext}
  \begin{equation}
  \label{eq:Lag_nuclear}
  \mathcal{L}_{\nu-N}=\sum_{X=\text{All}}\overline{\nu}f_X\Gamma_X\nu\,X
  + \sum_{X=S,V,T}\overline{N}\,\overline{C}_X\Gamma_X N\,X
  %\nonumber\\
  %&
  +\sum_{\substack{(X,Y)=(P,S)\\\;\;\qquad(A,V)}}\overline{N}\,i\overline{D}_X\Gamma_Y N\,X \ .
\end{equation}
\end{widetext}
Expressions for the coupling of the nucleus to the corresponding
mediator are given by \cite{AristizabalSierra:2018eqm}
\begin{align}
  \label{eq:nuclear-couplings}
  \overline{C}_S&=Z\sum_\mathrm{q}\frac{m_p}{m_\mathrm{q}}f_{T_\mathrm{q}}^p g^\mathrm{q}_S + (A-Z)\sum_\mathrm{q}\frac{m_n}{m_\mathrm{q}}f_{T_\mathrm{q}}^n g^\mathrm{q}_S\ ,
       \\
  \overline{C}_V&=Z(2 g^u_V + g^d_V) + (A-Z)(g^u_V + 2 g^d_V)\ ,
       \\
  \overline{C}_T&=Z(\delta_u^p g^u_T + \delta_d^p g^d_T) + (A-Z)(\delta_u^n g^u_T + \delta_d^n g^d_T)\ ,
\end{align}
where the different nucleon coefficients are obtained from chiral
perturbation theory from measurements of the $\pi$-nucleon sigma term
and from data of azimuthal asymmetries in semi-inclusive
deep-inelastic-scattering and $e^+ e^-$ collisions
\cite{Cheng:1988im,Anselmino:2008jk,Courtoy:2015haa,Goldstein:2014aja,Radici:2015mwa}. Expressions
for $D_P$ and $D_A$ can be obtained by replacing $g^\mathrm{q}_S\to h_P^\mathrm{q}$ and
$g_V^\mathrm{q}\to h_A^\mathrm{q}$ in $C_S$ and $C_V$, respectively.

The differential cross section induced by the simultaneous presence of
all the interactions in Eq.~(\ref{eq:Lag_nuclear}) can be adapted to
the light mediator case from the result derived in the effective limit
in Refs. \cite{Lindner:2016wff,AristizabalSierra:2018eqm}
\begin{widetext}
\begin{equation}
  \left.\frac{d\sigma}{dE_r}\right|_\text{NGI}=\frac{G_F^2}{2\pi}m_NF^2(q^2)
  \left[
  \xi_S^2\frac{2 E_r}{E_r^\text{max}} + \xi_V^2\left(2 - \frac{2 E_r}{E_r^\text{max}}\right)
  + \xi_T^2\left(2 - \frac{E_r}{E_r^\text{max}}\right)
  \right]\ .
\end{equation}
\end{widetext}
Here $E_r^\text{max}\simeq 2E_\nu^2/m_N$ and, in contrast to the
effective case, the $\xi_X$ parameters are $q^2=2m_NE_r$ dependent,
though they follow the same definitions 
\begin{equation}
  \label{eq:xi_parameters}
  \xi_S^2=C_S^2 + D_P^2\ , ~~ \xi_V^2=C_V^2 + D_A^2\ ,
   ~~ \xi_T^2= 4 C_T^2 \ . 
\end{equation}
The parameters in the right-hand side are in turn defined as: 
\begin{equation}
  \label{eq:cross_sec_parameters}
  C_X=\frac{1}{\sqrt{2}G_F}\frac{f_X\overline{C}_X}{2m_NE_r + m_X^2}\ ,
  \quad
  D_X=\frac{1}{\sqrt{2}G_F}\frac{f_X\overline{D}_X}{2m_NE_r + m_X^2}\ ,
\end{equation}
with the exception of $C_V$ which is shifted by the SM contribution,
$C_V\to Q_W + C_V$, with $Q_W$ given by
Eq.~(\ref{eq:weak-cahrge}). Two relevant remarks follow from the
expressions in Eqs.~(\ref{eq:xi_parameters}) and
(\ref{eq:cross_sec_parameters}). First of all, one can notice that in
the low momentum transfer limit and with $m_X\ll q$ the $\xi_X$
parameters are enhanced. This is at the origin of the spectral
distortions that could be expected if any of these interactions sneaks
in the signal. Secondly, unlike the effective case, where each $\xi_X$ can be treated as a free parameter
(thus allowing to encapsulate various
interactions at the same time, e.g. in $\xi_S$ a scalar and
pseudoscalar interaction), in this case the $q^2$ dependence does not allow
that. Thus, if one considers e.g. $\xi_S$, in full generality a
four-parameter analysis is required. To assess the impact of the
Dresden-II reactor experiment signal, we then proceed by assuming a
single mediator at a time: $\xi_S$ determined only by $C_S$ and
$\xi_V$ by $Q_W+C_V$. 
Let us finally note that for the case of $\xi_T$, such an assumption is not necessary.
  
\begin{figure*}[t]
  \centering
  \includegraphics[scale=0.35]{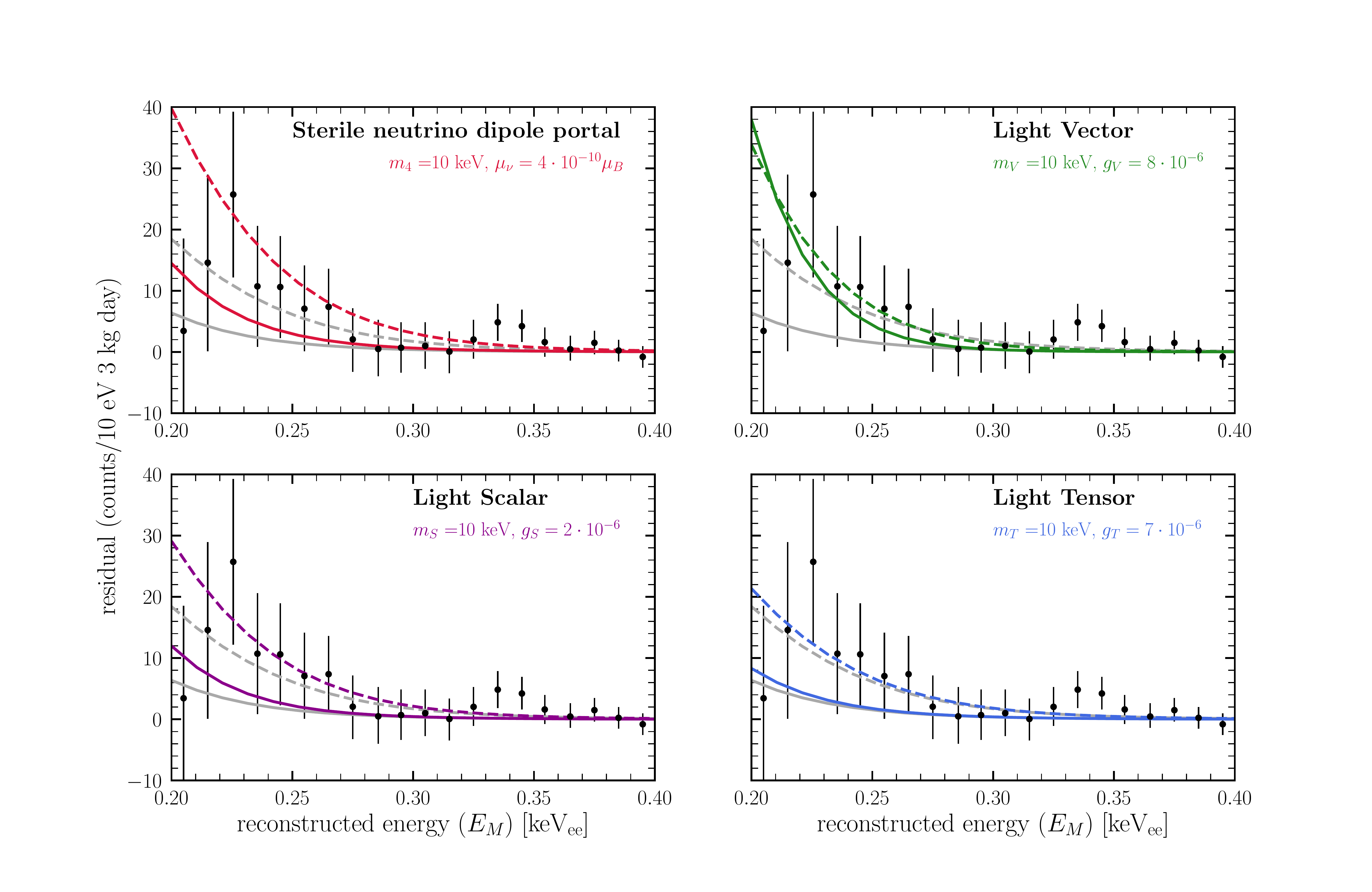}
  \caption{Experimental data from the Dresden-II reactor obtained
    during 96.4 days exposure time using the NNC-1701 germanium
    detector. CE$\nu$NS data follow from residual counts after
    the subtraction of the best-fit background components
    \cite{Colaresi:2022obx}. The spectral rates of signal events
    are also shown, for the SM prediction obtained with the modified Lindhard QF [see
    Eq.~(\ref{eq:QF_Linhard_modified})] (gray curves, solid for $q=0$
    and dashed for $q = -20\times 10^{-5}$, in both cases $k=0.157$)
    and for various new physics scenarios, with same assumptions on the
    QF.}
  \label{fig:data-points}
\end{figure*}
% -------------
% Section
% -------------

\subsection{Sterile neutrino dipole portal}
\label{sec:nu-F-magnetic-transition}
In the Dirac case neutrino magnetic and electric dipole moment
couplings are dictated by the following Lagrangian
\cite{Grimus:2000tq}

\begin{equation}
  \label{eq:Lag_mag_moment}
  \mathcal{L}=\overline{\nu}\,\sigma_{\mu\nu}\,\lambda\,\nu_R\,F^{\mu\nu}
  +
  \text{H.c.}\ ,
\end{equation}
where in general $\lambda$ is a $3\times N$ matrix in flavor
space. These couplings are chirality flipping and so the scattering
process induced by an ingoing active neutrino produces a sterile
neutrino in the final state. Thus, Dirac neutrino magnetic moments
always induce up-scattering processes ($\nu_L+N\to F_4+N$). The mass
of the outgoing fermion, being a free parameter, is only
constrained by kinematic criteria. Given an ingoing neutrino energy
$E_\nu$, its mass obeys the following relation:
\begin{equation}
  \label{eq:kinematic_constraint}
  m_4^2\lesssim 2m_N E_r\left(\sqrt{\frac{2}{m_N E_r}}E_\nu -1\right)\ .
\end{equation}
For the nuclear recoil energies involved at  the Dresden-II experiment and for
neutrino energies near the kinematic threshold, $E_\nu \sim 9.5\,$MeV, the upper
bound $m_4\lesssim 8\,$MeV applies. 

The interactions in Eq.~(\ref{eq:Lag_mag_moment}) contribute to
the CE$\nu$NS cross section \cite{McKeen:2010rx} through
\begin{widetext}
\begin{equation}
  \label{eq:dipole_portal_xsec}
  \left. \frac{d\sigma}{dE_r}\right|_\text{DP}=
  \alpha_\text{EM}\,\mu_{\nu,\text{Eff}}^2\,F^2(q^2) Z^2\,
  \left[\frac{1}{E_r} - \frac{1}{E_\nu} 
    - \frac{m_4^2}{2E_\nu E_rm_N}
    \left(1- \frac{E_r}{2E_\nu} + \frac{m_N}{2E_\nu}\right)
    + \frac{m_4^4(E_r-m_N)}{8E_\nu^2E_r^2m_N^2}
  \right]\ .
\end{equation}
\end{widetext}
Here $\alpha_\text{EM}$ refers to the electromagnetic fine structure
constant and $\mu_{\nu,\text{Eff}}$ to a dimensionless [normalized to
the Bohr magneton, $\mu_B=e/(2m_e)$] parameter space function that
involves combinations of the entries of the $\lambda$ matrix weighted
by neutrino mixing angles and possible CP phases (for details see
 \cite{AristizabalSierra:2021fuc,Miranda:2021kre}). Note that in the limit
$m_4\to 0$, Eq.~(\ref{eq:dipole_portal_xsec}) matches the ``standard''
neutrino magnetic moment cross section \cite{Vogel:1989iv}.
% --------------
% Section
% --------------
\section{The data, the recoil spectrum 
  and the statistical analysis}
\label{sec:parameter_space_analysis}
In this section we present a brief discussion of the data reported by
the Dresden-II reactor experiment, provide the technical tools that
allow the calculation of the CE$\nu$NS signal (within the SM and with
new physics) and present our statistical analysis along with our
results for the scenarios discussed in Sec. \ref{sec:ngi-lm}.
\subsection{Data and recoil spectra}
\label{sec:data_plus_spectra}
The Dresden-II reactor experiment consists of a p-type point contact
(PPC) 2.924 kg ultra-low noise and low energy threshold (0.2
keV$_{\rm ee}$) germanium detector located at $\sim 10\,$m from the
2.96 GW Dresden-II boiling water reactor (BWR): The NCC-1701 detector
\cite{Colaresi:2021kus}. The proximity to the detector along with its
high power implies a high flux of electron anti-neutrinos.  The data
accumulated during 96.4 days of effective exposure with the reactor
operating at nominal power (Rx-ON), hint to a first ever observation
of CE$\nu$NS using reactor neutrinos, as recently reported in
Ref. \cite{Colaresi:2022obx}.  The residual difference between the
full spectrum and the best-fit background components (the suggested
CE$\nu$NS signal) spans over the measured energy range
$E_M\subset [0.2,0.4]\,$keV$_\text{ee}$ and involves 20 data bins
equally spaced ($0.01\,$keV$_\text{ee}$), as shown in
Fig.~\ref{fig:data-points}.

The CE$\nu$NS differential recoil energy spectrum follows from a
convolution of the electron anti-neutrino flux and the CE$\nu$NS cross
section, namely
\begin{equation}
  \label{eq:recoil-spectrum}
  \frac{dR}{dE_r}=N_T\int_{E_\nu^\text{min}}^{E_\nu^\text{max}}
  \,\frac{d\Phi_{\overline{\nu}_e}}
  {dE_\nu}\frac{d\sigma_\text{CE$\nu$NS}}{dE_r}\,dE_\nu\ .
\end{equation}
The number of germanium nuclei in the detector is given by
$N_T=m_\text{det}\,N_A/m_\text{Ge}$, with $N_A$ the Avogadro number,
$m_\text{Ge}$ the germanium molar mass and
$m_\text{det}=2.924\,$kg. The integration limits are given by
$E_\nu^\text{min}=\sqrt{m_N E_r/2}$, with $E_r$ being the recoil energy, and
$E_\nu^\text{max}$ the kinematic value determined by the electron
anti-neutrino flux. We take the
values of the atomic number and nuclear mass for
 $^{72}\text{Ge}$. For neutrino energies below 2 MeV we use
the anti-neutrino spectral function from Kopeikin
\cite{Kopeikin:2012zz}, while for energies above that value we consider
Mueller et al.~\cite{Mueller:2011nm}. For flux normalization we use
$\mathcal{N}=4.8\times
10^{13}\overline{\nu}_e/\text{cm}^2/\text{sec}$,
as given in Ref.~\cite{Colaresi:2022obx}. The differential
anti-neutrino flux in Eq.~(\ref{eq:recoil-spectrum}) therefore involves the
spectral function and the normalization. The CE$\nu$NS
differential cross section is dictated by Eq.~(\ref{eq:xsec_CEvNS}),
but can also involve contributions from NGI couplings or the sterile
neutrino dipole portal discussed in Sec. \ref{sec:renormalizable-NGI}
and \ref{sec:nu-F-magnetic-transition}.

\begin{figure*}
  \centering
  \includegraphics[scale=0.43]{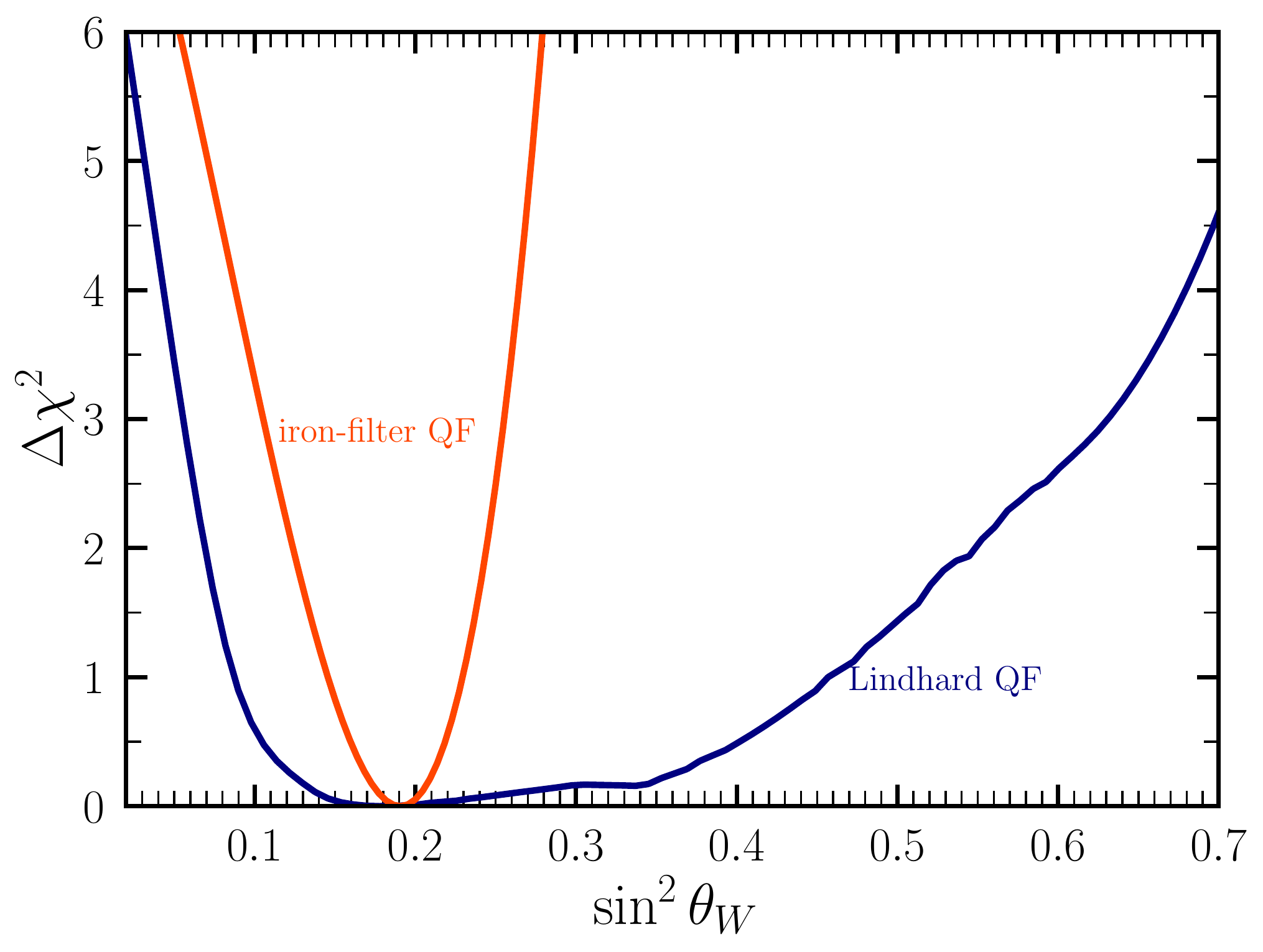}
  \caption{$\Delta\chi^2$ profiles for $\sin^2\theta_W$ for the two
    QFs considered [modified Lindhard QF,
    Eq.~(\ref{eq:QF_Linhard_modified}), and iron-filter QF as given in
    the ancillary files in Ref. \cite{Colaresi:2022obx}]. For the
    modified Lindhard QF the result follows after marginalization over
    $k$ and $q$ [see Eq.~(\ref{eq:QF_Linhard_modified})]. }
  \label{fig:weak-mixin-angle-chiSq-single-parameter}
\end{figure*}
For detectors relying on ionization (it applies to scintillation as
well), such as the NCC-1701, only a fraction of the nuclear recoil
energy is available in a readable format. The characterization of that
fraction is given by the QF, $Q$, defined as the ratio between the nuclear
recoil given in ionization ($E_I$) over that generated by an electron
recoil of the same kinetic energy ($E_r$). Quantitatively, this means
that the ionization energy expected from a given recoil energy is
given by $E_I=Q\,E_r$. With the aid of the QF, the differential
ionization spectrum can then be written according to
\begin{equation}
  \label{eq:ionization-spectrum}
  \frac{dR}{dE_I}=\frac{dR}{dE_r}\frac{dE_r}{dE_I}=
  \frac{dR}{dE_r}\left(\frac{1}{Q}-\frac{E_I}{Q^2}\frac{dQ}{dE_I}\right)\ .
\end{equation}
For sufficiently high-recoil energy regimes (above 5 keV$_\text{nr}$
or so) the QF is well described by the Lindhard model
\cite{osti_4701226}. However, its validity is
questionable in any material for sub-keV energies, as pointed out in
Ref. \cite{osti_4153115}. For germanium, recent measurements of its QF
using recoils from gamma emission following thermal neutron capture,
photo-neutron sources, and a monochromatic filtered neutron beam have
shown substantial deviations from the Lindhard model expectations at
recoil energies below $\sim 1.3\,\text{keV}_\text{nr}$
\cite{Collar:2021fcl}.  In the context of DM direct detection
searches, Ref. \cite{Sorensen:2014sla} has addressed this issue
providing a slight modification of the Lindhard QF
\begin{equation}
  \label{eq:QF_Linhard_modified}
  Q(E_r)=\frac{k\,g(\epsilon)}{1 + k\,g(\epsilon)}-\frac{q}{\epsilon}\ ,\\[4mm]
\end{equation}
where the first term is the standard Lindhard QF with
$g(\epsilon)=3\epsilon^{0.15}+0.7\epsilon^{0.6}+\epsilon$ and
$\epsilon=11.5 Z^{-7/3}\,E_r$. The second term (the correction) is
such that deviations from the standard behavior start to show up at
about $0.1\,$keV. In our analyses we adopt this parametrization, and
therefore we include $k$ and $q$ as free parameters.  In addition to
this QF, we employ as well the ``iron-filter'' QF reported in the
ancillary files provided by Ref. \cite{Colaresi:2022obx}.

The CE$\nu$NS ionization differential spectrum in
Eq.~(\ref{eq:ionization-spectrum}) has to be smeared by the intrinsic
resolution of the detector. Following the information of the {\tt
  README} ancillary file \cite{Colaresi:2022obx}, we take the resolution to be a Gaussian
truncated energy-dependent distribution given by~\cite{Coloma:2022avw}
\begin{equation}
  \label{eq:gaussian}
  G(E_M,E_I,\sigma)=\frac{2}{1+\text{erf}(E_I/\sqrt{2}/\sigma)}
  \frac{1}{\sqrt{2\pi}\sigma}e^{-\Delta E^2/2/\sigma^2}\ .
\end{equation}
Here, the energy-dependent Gaussian width $\sigma^2=\sigma_n^2+E_I\,\eta\,F$ involves
the intrinsic electronic noise of the detector $\sigma_n=68.5\,$eV
(for the 96.4 days of Rx-ON data), the average energy of $e^-$-hole
formation in germanium $\eta=2.96\,$eV, and the Fano factor whose
value we fix to the average value in the range [0.10-0.11],
$F=0.105$. As stressed in the ancillary file, overall the second term
in the Gaussian width measures the dispersion in the number of
information carriers ($e^-$-hole pairs).

The smearing of the ionization differential spectrum results in the
measured differential spectrum
\begin{equation}
  \label{eq:measured_diff_spectrum}
  \frac{dR}{dE_M}=\int_{\eta}^\infty\,G(E_M,E_I,\sigma)\,\frac{dR}{dE_I}\,dE_I\ ,
\end{equation}
from which the number of events in the $i$th bin is obtained by
integration over the measured energy $E_M$, in the interval
$[E_M^i-\Delta E_M,E_M^i+\Delta E_M]$
($\Delta E_M=5\,\text{eV}_\text{ee}$). The integration lower limit is set by the minimum average ionization energy $\eta \sim 3~\mathrm{eV_{ee}}$ required to produce
an $e^-$-hole pair in germanium.

\subsection{Statistical analysis}
\label{sec:statistics}
Our analysis is based on the $\chi^2$ function
\begin{equation}
  \label{eq:chiSq_dist}
  \chi^2 (\vec{S},\alpha) =  \sum_i \left[
 \frac{N_\text{th}^{i}(\vec{S}, \alpha) - N_\text{meas}^{i}}{\sigma_i} \right]^2
  + 
  \left( \frac{\alpha}{\sigma_{\alpha}}\right)^2 \, ,
\end{equation}
where $N_\text{th}^{i}$ and $N_\text{meas}^{i}$ are the theoretical
and measured number of events, respectively, in the $i$th energy
bin. Note that in the definition of the $\chi^2$ function we are
assuming the data to follow a Gaussian distribution. Although assuming
a Poisson distribution would be a better choice given the dataset,
both statistical and systematic errors (which have a bigger impact on
the results) can be readily included under the Gaussian
assumption. Here, $\sigma_i$ represents the corresponding uncertainty
of the $i$th measurement which includes systematic and statistical
uncertainties. Here, $\vec{S}$ represents a set of new physics
parameters, while $\alpha$ is a nuisance parameter which accounts for
the flux normalization uncertainty, for which we consider
$\sigma_{\alpha} = 5\%$. The theoretical number of events is
\begin{equation}
  \label{eq:theoretical_number_events}
 N_\text{th}(\vec{S}, \alpha)=  (1+\alpha) N_\text{CEvNS}(\vec{S})\ ,
\end{equation}
which, of course, includes the SM piece in addition to the new physics contribution.

Equipped with the tools discussed in
Sec. \ref{sec:parameter_space_analysis} along with the $\chi^2$
function in Eq.~(\ref{eq:chiSq_dist}), we begin our discussion by focusing on
the implications for the weak mixing
angle. Figure~\ref{fig:weak-mixin-angle-chiSq-single-parameter} shows
the $\Delta\chi^2$ distributions in terms of $\sin^2\theta_W$ for the two
QFs considered in the analysis. In the case of the modified Lindhard
QF, our result is obtained by marginalizing over the parameters $k$
and $q$ [see Eq.~(\ref{eq:QF_Linhard_modified})]. Notice that the $\Delta \chi^2$ profile for the case of Lindhard QF is rather flat at the bottom, thus making its best fit value not very statistically meaningful. Specifically, the Lindhard parameters are allowed to float in the ranges\footnote{Note, that $k=0.27$ corresponds to the limit set by CONUS~\cite{Bonet:2020awv}.} $0.14 \leq k \leq 0.27$ and $-40 \leq q/10^{-5} \leq 0$. As expected, a
strong dependence on the QF is observed. The best-fit values differ by
about $~\sim 6.5 \%$, with the iron-filter QF favoring a larger
$\sin^2\theta_W$ value. The $1\sigma$ ranges read
\begin{alignat}{2}
  \label{weak_mix_angle_1_sigma}
  \text{Modified Lindhard QF:}&\quad&\sin^2\theta_W&=0.178^{+0.280}_{-0.090}\,
  \nonumber\\
  \text{Iron filter QF:}&\quad&\sin^2\theta_W&=0.190^{+0.039}_{-0.046}\ ,
\end{alignat}
thus showing the disparity of the values obtained as a consequence of
a different QF model.  One can notice as well that both values differ
substantially from the SM RGE expectation. In particular, the best fit result from the iron-filter QF analysis is compatible with the SM RGE prediction
at $80.7\%$ C.L., whereas the result from the modified Lindhard QF is in agreement at $1\sigma$, 
given the spread of its $\Delta \chi^2$ distribution.
From these results one can conclude that with the current data
set and the lack of a better knowledge of the germanium QF, a robust
determination of the weak mixing angle seems not possible.

\begin{figure*}
  \centering
  \includegraphics[scale=0.45]{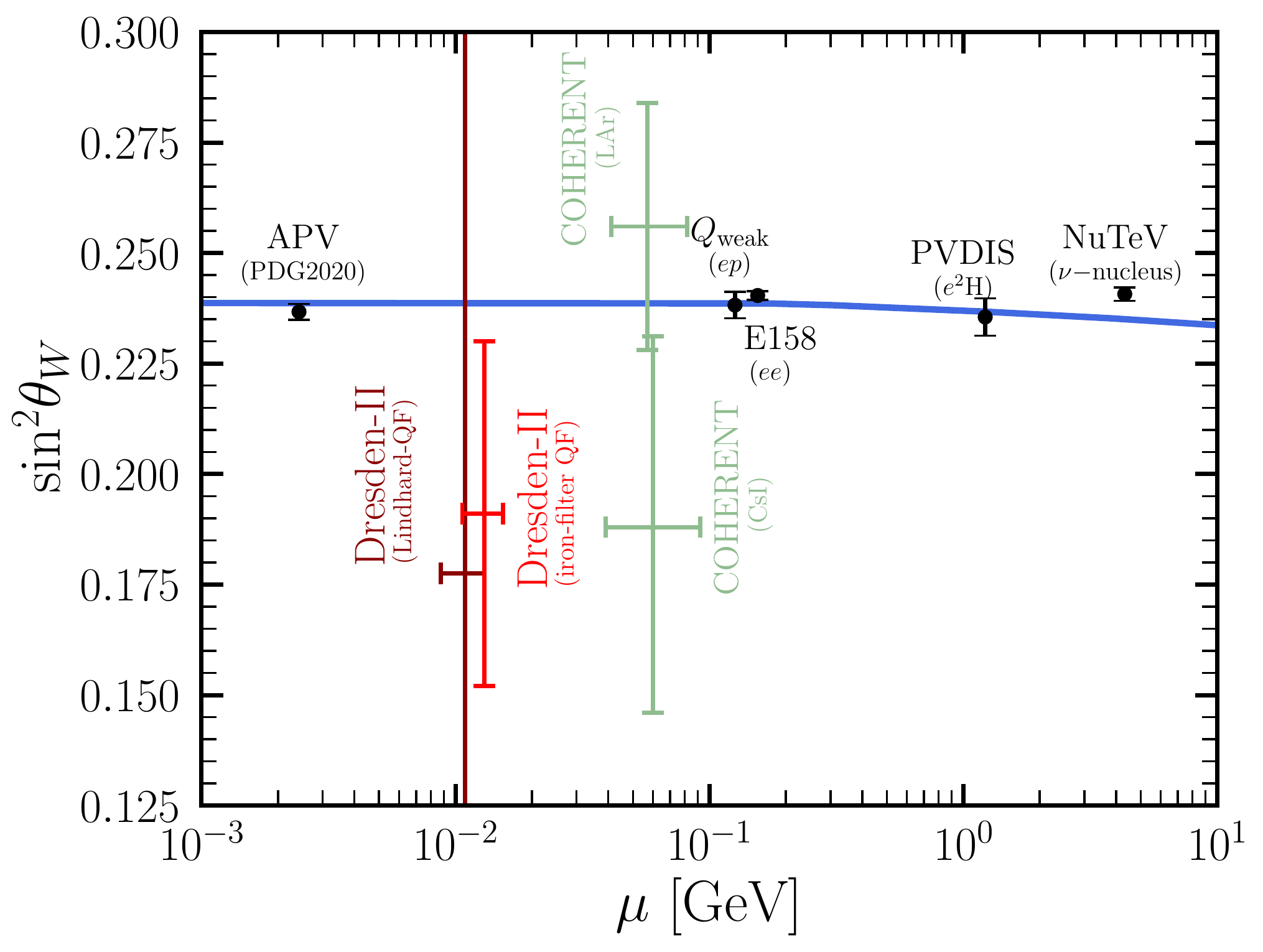}
  \caption{Weak mixing angle RGE running in the SM, calculated in the
    $\overline{\text{MS}}$ renormalization scheme as obtained in
    Ref. \cite{Erler:2004in}, along with measurements at different
    renormalization scales
    \cite{Akimov:2017ade,COHERENT:2020iec,Wood:1997zq,Derevianko:2010kz,Androic:2018kni,Anthony:2005pm,Wang:2014bba,Zeller:2001hh}. The
    $1\sigma$ result obtained using the Dresden-II reactor data is
    shown assuming the modified Lindhard and iron-filter QF (see text for further details).}
  \label{fig:RGE_evolution}
\end{figure*}

Although featuring a moderate disparity, these results can be understood as
a first determination of the weak mixing angle at low energies using
CE$\nu$NS data from reactor anti-neutrinos. They can be compared with
the values obtained from COHERENT CsI and LAr data
\cite{Akimov:2017ade,COHERENT:2020iec} and other dedicated experiments
that include atomic parity violation (APV)
\cite{Wood:1997zq,Derevianko:2010kz}, proton weak charge from Cs
transitions ($Q_\text{weak}$) \cite{Androic:2018kni}, M{\o}ller
scattering (E158) \cite{Anthony:2005pm}, parity violation in deep
inelastic scattering (PVDIS) \cite{Wang:2014bba} and neutrino-nucleus
scattering (NuTeV) \cite{Zeller:2001hh}. A summary of these results is
displayed in Fig.~\ref{fig:RGE_evolution}, which shows as well the RGE
running calculated in the $\overline{\text{MS}}$ renormalization
scheme \cite{Erler:2004in}. The value for the weak mixing angle at the
$1\sigma$ level extracted from the best fit in
Fig.~\ref{fig:weak-mixin-angle-chiSq-single-parameter} is shown. For the renormalization scale at which the
measurement applies, we have adopted a rather simple procedure. We
have translated the ionization energy range into recoil energy with
the aid of the QF.  With the values obtained for $E_r^\text{min}$ and
$E_r^\text{max}$ we have then calculated the momentum transfer by
using the kinematic relation $q^2=2m_NE_r$. This result corresponds to the first CE$\nu$NS-based determination of $\sin^2\theta_W$ with reactor data at
$\mu \sim 10\,$MeV.  With further data, and more importantly a
better understanding of the germanium QF, this result is expected to
highly improve in the future.

We now move on to the case of NGI. For this analysis we assume
universal quark couplings and switch off the pseudovector couplings in
the vector case (those controlled by $\xi_V$) as well as the pseudoscalar
couplings in the scalar case (those controlled by $\xi_S$). These
simplifications reduce the analysis to pure vector and pure scalar
interactions, controlled by the couplings $g_V^2 = g_V^\mathrm{q} f_V$
and $g_S^2 = g_S^\mathrm{q} f_S$ (and the mediators masses), as
investigated in Refs. \cite{Liao:2022hno,Coloma:2022avw}. For the
tensor case no assumption on different contributions is required. The
cross section is determined by $\xi_T$ and, under the assumption of
universal quark couplings, it is eventually controlled by
$g_T^2 = g_T^\mathrm{q} f_T$. Again, for the statistical analysis using the
modified Lindhard QF we vary as well $q$ and $k$.  The analysis in
this case is therefore a four parameter problem, while for the
iron-filter QF only two parameters matter, i.e. the new mediator mass and
coupling.

\begin{figure*}
  \centering
  \includegraphics[scale=0.4]{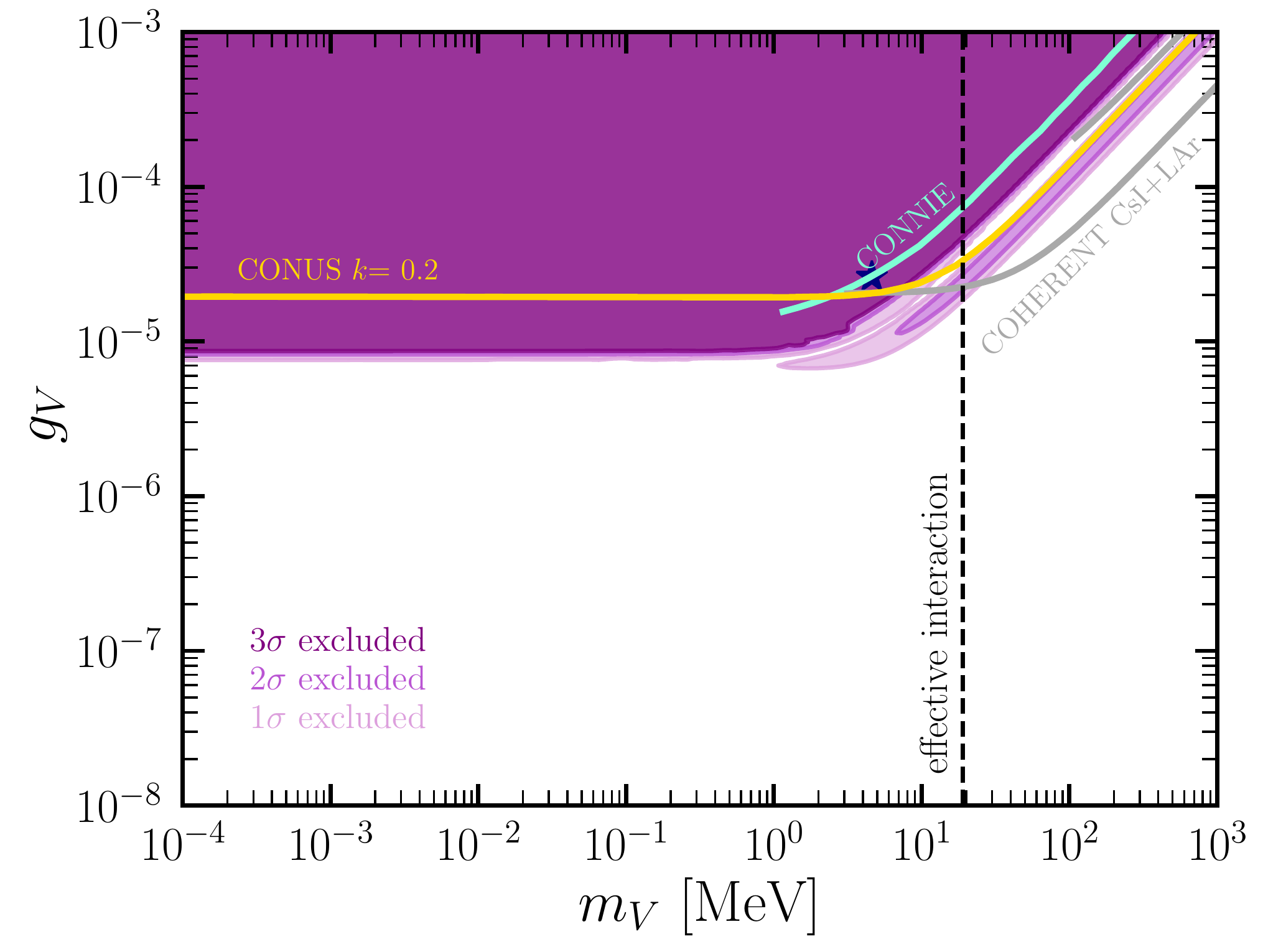}
  \includegraphics[scale=0.4]{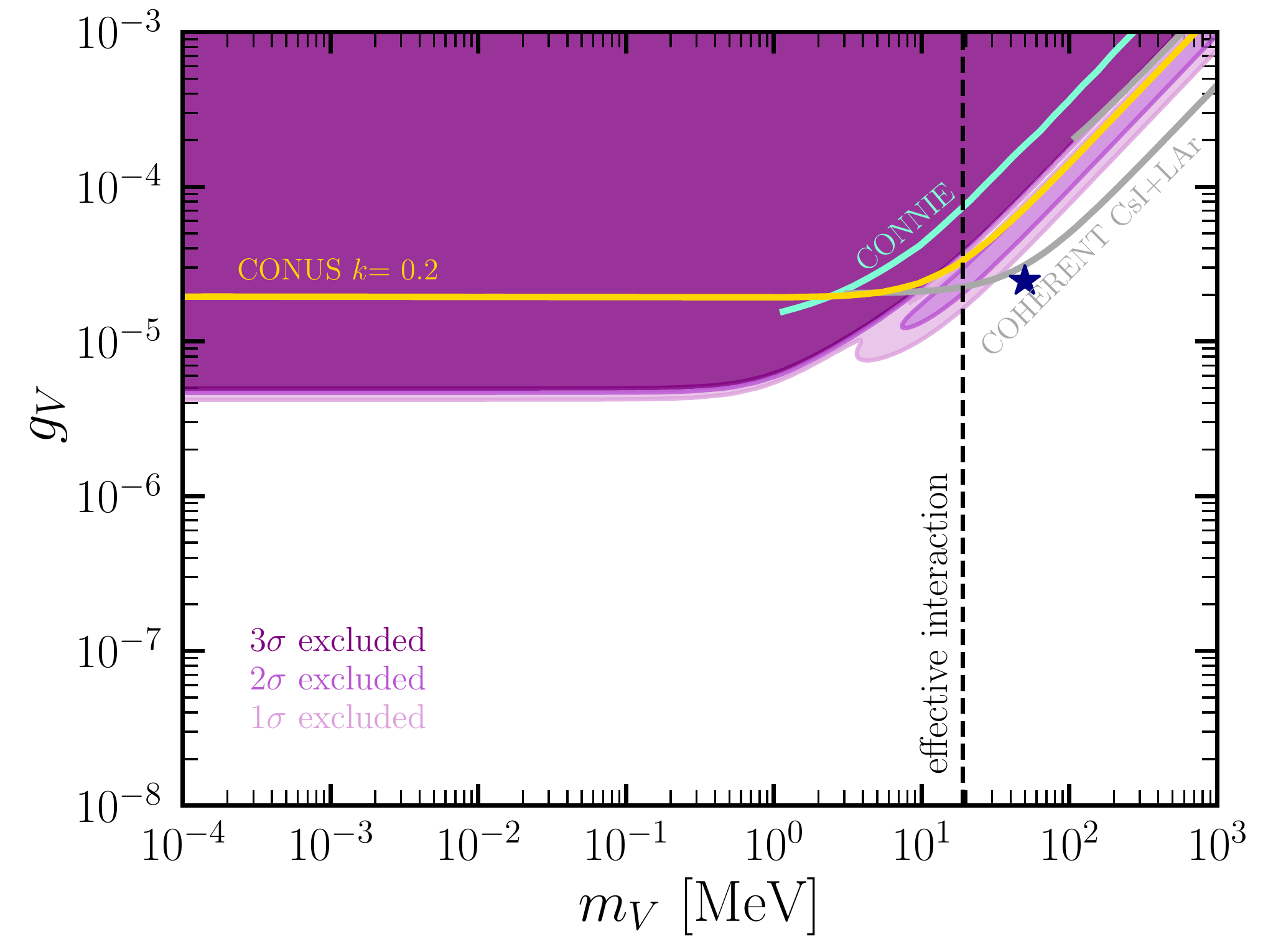}
  
  \includegraphics[scale=0.4]{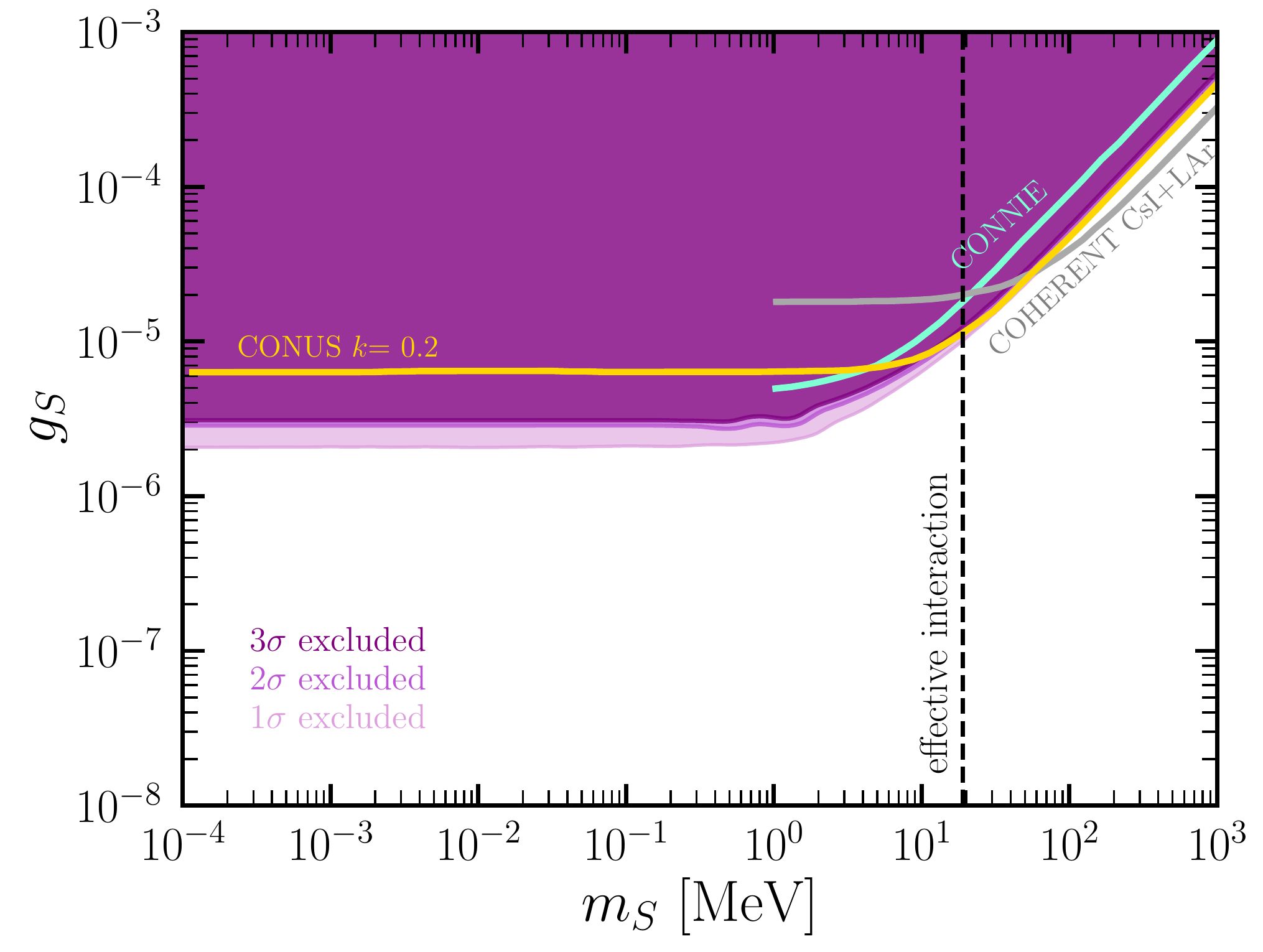}
  \includegraphics[scale=0.4]{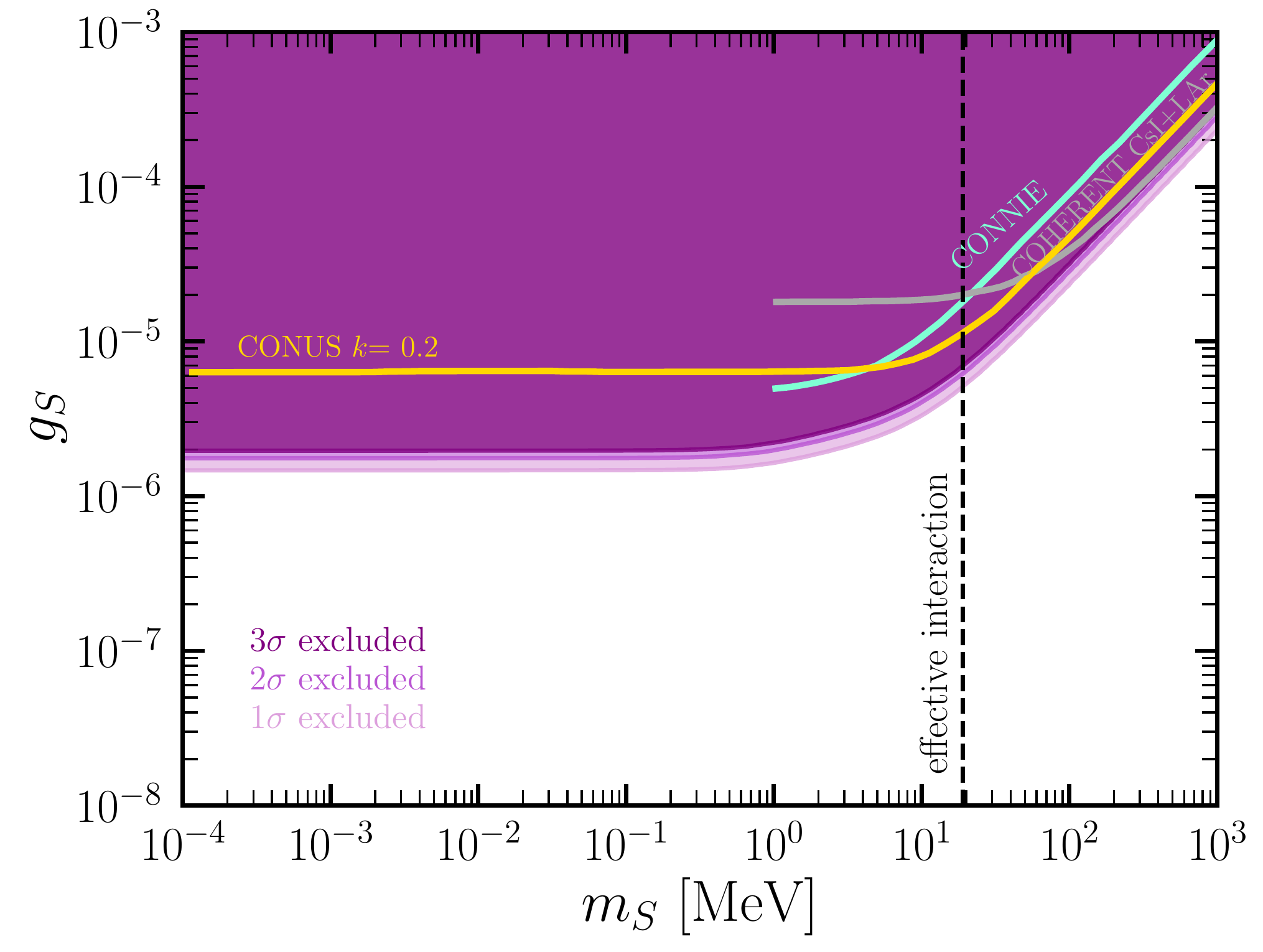}
  
  \includegraphics[scale=0.4]{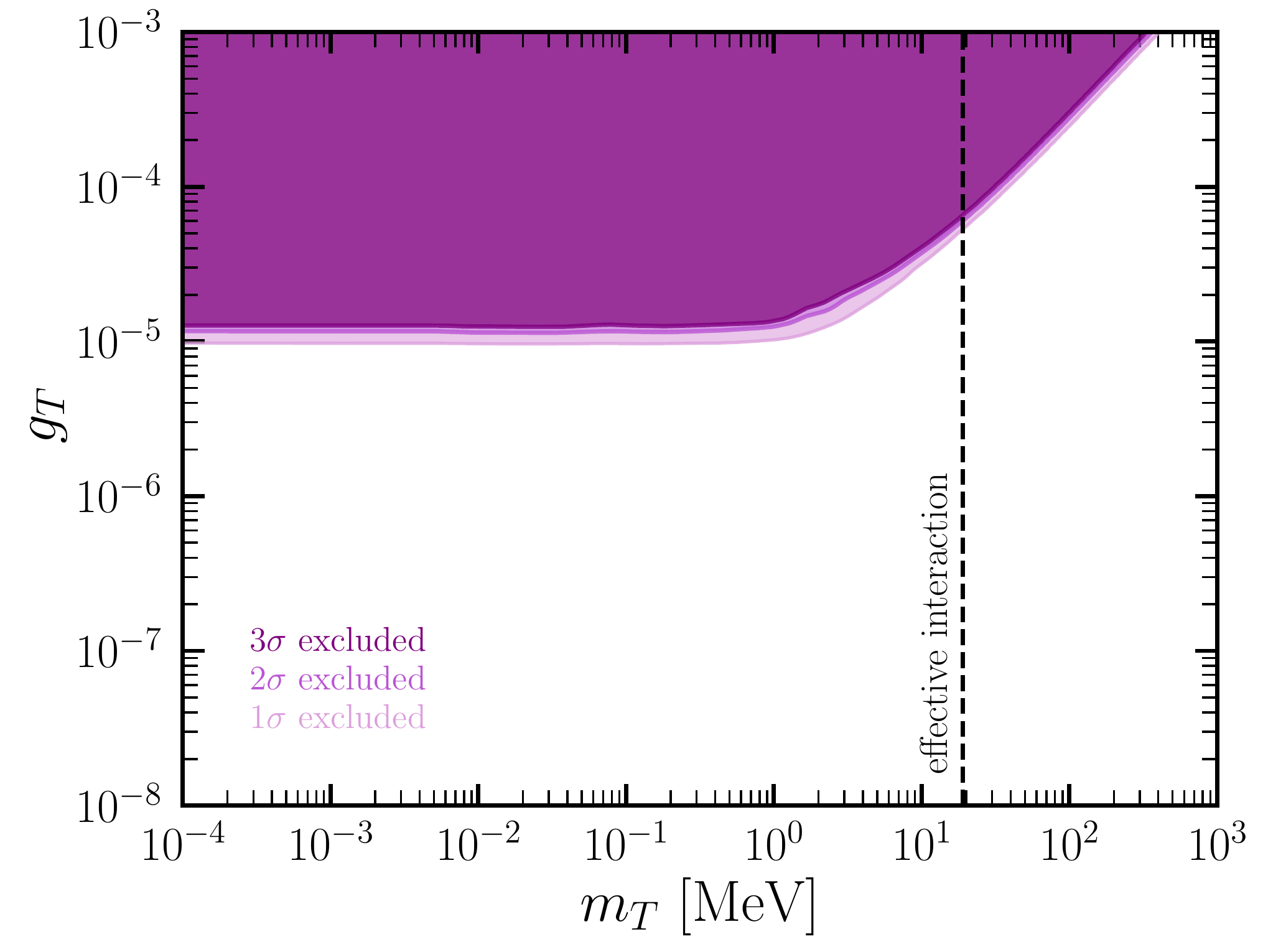}
  \includegraphics[scale=0.4]{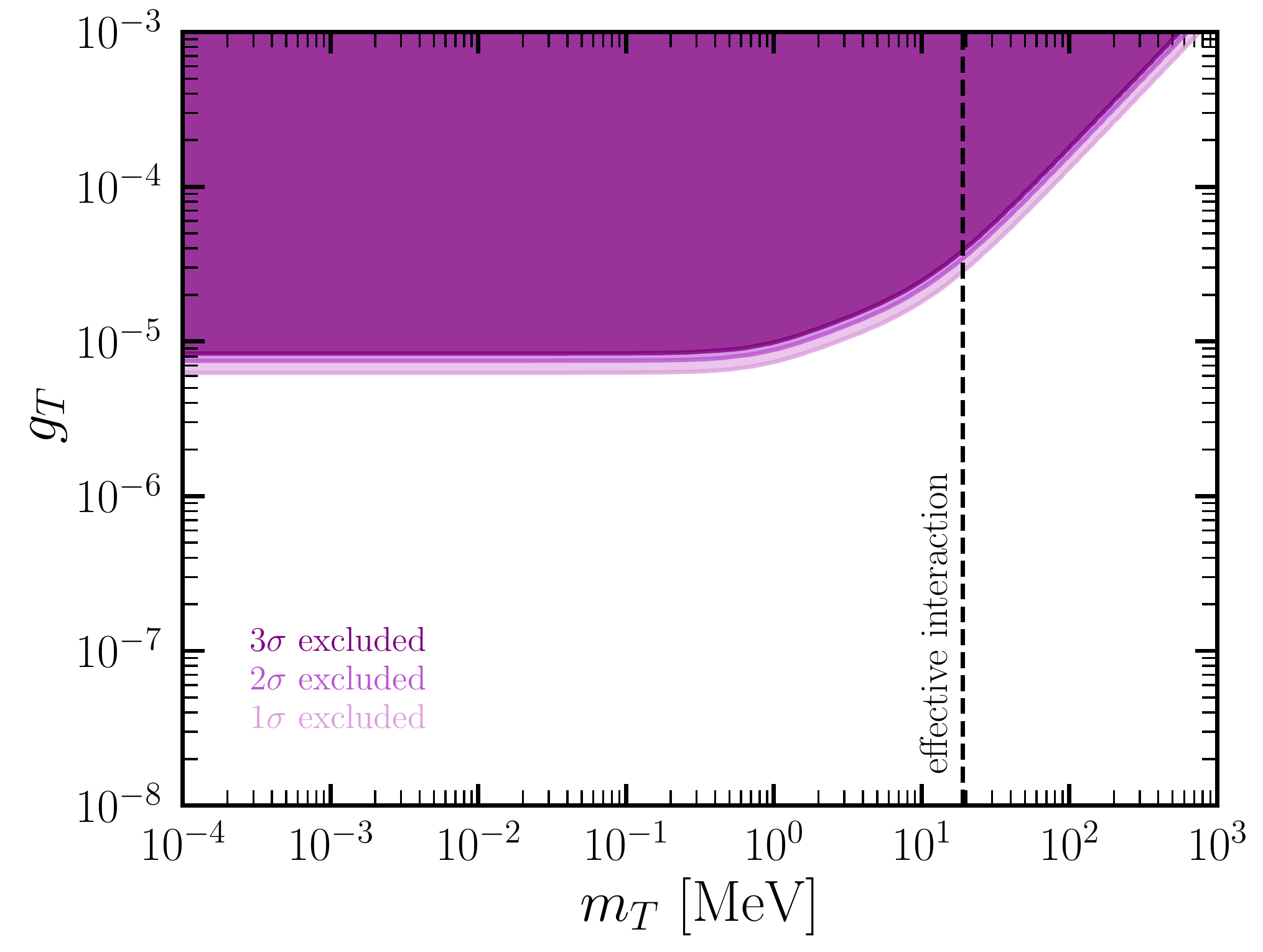}
  \caption{Constraints on vector NGI (upper row), scalar NGI (central
    row) and tensor NGI (lower row) in the coupling-mass plane,
    obtained using the modified Lindhard QF (left column) and the
    iron-filter QF (right column).  In all panels, purple regions
    indicate exclusion limits.  Where present, dark blue stars specify
    the best fit solutions.  Moreover, constraints from
    CONUS~\cite{CONUS:2021dwh}, CONNIE~\cite{CONNIE:2019xid} and
    COHERENT CsI+LAr~\cite{Corona:2022wlb} are shown for
    comparison. Additionally, the black dashed vertical line marks the
    transition from the light to the effective regime.}
  \label{fig:NGI}
\end{figure*}
Our extracted sensitivities are illustrated in Fig.~\ref{fig:NGI} at $1,2,3~\sigma$ (assuming two d.o.f., i.e. $\Delta \chi^2 = 2.3, 6.18, 11.83$ respectively). The
upper row stands for the vector case, the middle row for the scalar
and the bottom row for the tensor, while left (right) panels are
obtained using the modified Lindhard (iron-filter) QF.  As can be
seen, at the $1\sigma$ level and above, large portions of parameter
space are ruled out, disfavoring couplings as low as $7.5\times 10^{-6}$
for $m_V \lesssim 100$ keV. At the $1$ and $2\sigma$ level, two
``islands'' in the region of noneffective interactions
($m_V\gtrsim 10\,$MeV) are open as well. At the $3\sigma$ level these
spots are gone and the constraint becomes a little less
stringent. Turning to the analysis done assuming the iron-filter QF,
we find that about the same regions in parameter space are excluded,
though the most stringent limit is a little more pronounced in this
case ($4\times 10^{-6}$ for $m_V\lesssim 100\,$keV).  The parameter
space ``islands'' found with the modified Lindhard QF are present in
this case as well, but cover a somewhat wider area. At the
$90 \%\,$C.L.  constraints on the vector NGI scenario amount to
$g_V\lesssim 8\times 10^{-6}$ (Lindhard QF) and
$g_V\lesssim 4.5\times 10^{-6}$ (iron-filter QF) for vector masses up
to 100 keV.  This limit should be compared with results from COHERENT
CsI and LAr, for which Refs. \cite{Liao:2017uzy,Miranda:2020tif} found
$g_V\lesssim 6\times 10^{-5}$ at the $90\%\,$C.L. We can then conclude
that the Dresden-II data largely improve limits for vector
interactions in the low vector mass window. This result can be
attributed to the sub-keV recoil energy threshold the experiment
operates with.

In the scalar case the situation is as follows. The modified Lindhard
QF and the scalar hypothesis tend to produce smaller deviations from
the data. This can be readily understood from the left graph in the
bottom row of Fig.~\ref{fig:data-points}. At low scintillation energy
the event rate tends to increase, but slightly less than in the vector
case, a behavior somehow expected, see
e.g. Ref. \cite{AristizabalSierra:2019ykk}. While the scalar coupling
contributes to the CE$\nu$NS cross section quadratically, the vector
does it linearly because of its interference with the SM
contribution. As a consequence, at $1\sigma$ level and above, limits
are slightly less stringent than in the vector case. In contrast to
that case as well, the parameter space ``islands'' are gone. Their
disappearance can be traced back to the fact that these interactions
do not sizably interfere with the SM term. Limits for scalar masses
below $\sim 1\,$MeV at the $90\%\,$C.L. amount to
$g_S\lesssim 3 \times 10^{-6}$ in the Lindhard QF case. For COHERENT
CsI and LAr, Refs. \cite{Papoulias:2017qdn,Miranda:2020tif} found
$g_S\lesssim 3.0\times 10^{-5}$ at the $90\%\,$C.L., implying a slight
improvement on the limit. For the iron-filter QF one finds about the
same trend, with limits at different statistical significances
spreading uniformly. The $90\%\,$C.L. limit at low scalar mass
amounts to $g_S\lesssim 1.8 \times 10^{-6}$, for scalar masses up to
$100\,$keV.

\begin{figure*}
  \centering
  \includegraphics[scale=0.4]{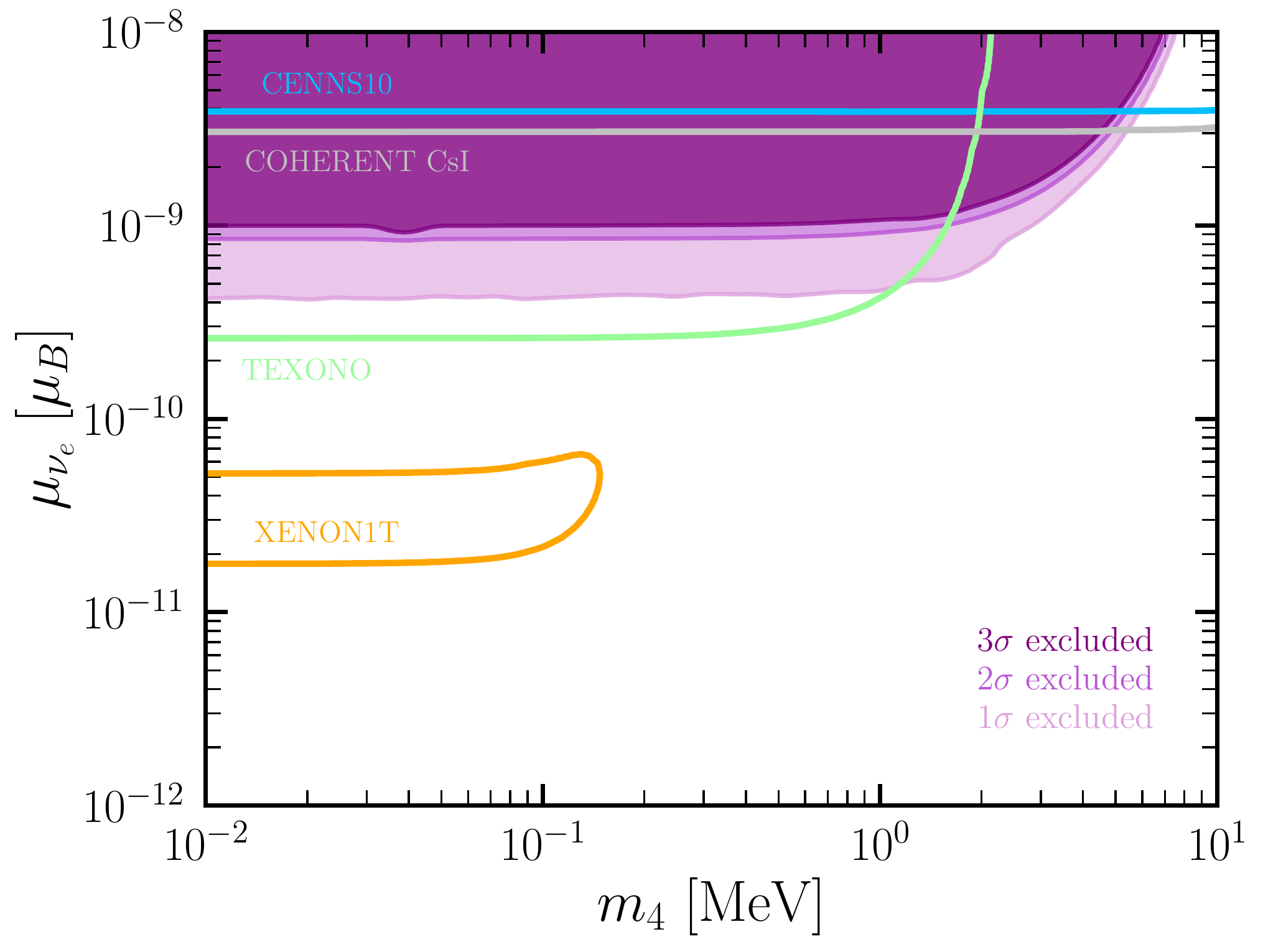}
  \includegraphics[scale=0.4]{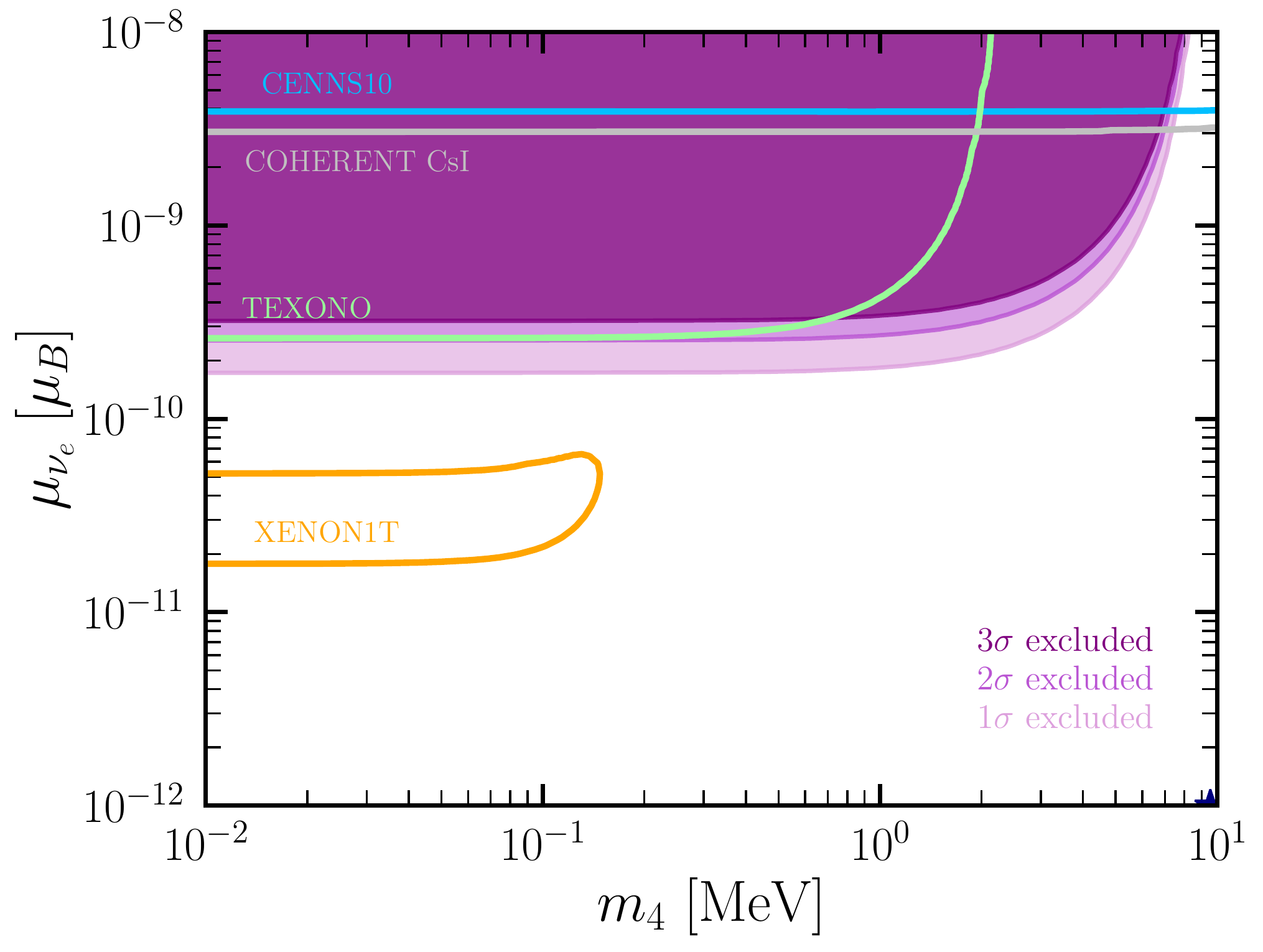}
  \caption{Results of the analysis for the sterile neutrino dipole
    portal based on two QF hypotheses: Modified Lindhard QF (left graph)
    and iron-filter QF (right graph). For the former case results
    follow after marginalization over $q$ and $k$. Shaded areas
    indicate the excluded regions at different statistical
    significance levels: $1\sigma$, $2\sigma$ and $3\sigma$ as shown
    in the graphs. Constraints from CENNS10, TEXONO, COHERENT CsI and
    XENON1T (see Ref.~\cite{Miranda:2021kre}) are also shown for
    comparison. }
  \label{fig:dipole-portal}
\end{figure*}
Results for the light tensor case resemble those found in the NGI
light scalar scenario, though limits are a little weaker. At the
$1\sigma$ level and above, we find
 $g_T\lesssim 1.0\times 10^{-5}$ (Lindhard QF) and $g_T\lesssim 6.0\times 10^{-6}$ (iron-filter QF) for tensor masses below
$\sim 100\,$keV.  Although with small differences, among the NGI we
have considered, the tensor couplings are the less constrained by the
Dresden-II data set. This result is inline with that found when
analyzing tensor NGI using CsI COHERENT data
\cite{AristizabalSierra:2018eqm}.

To our knowledge, limits on light tensor interactions using COHERENT
CsI and LAr data have been discussed only in \cite{Demirci:2021zci}.
On the other hand, there are some forecasts for searches for this type
of interactions at multi-ton DM detectors
\cite{Majumdar:2021vdw}. Searches relying on the CE$\nu$NS nuclear
recoil channel are expected to be sensitive up to
$g_T\sim 2.0\times 10^{-5}$ for tensor masses up to $\sim 1\,\text{MeV}$
at the $90\%\,$C.L. These numbers lead to the same conclusion than in
the scalar case: In the light mediator regime, constraints obtained
using Dresden-II data seem to improve upon available sensitivities.

As we have already pointed out, given the kinematic threshold of the
electron anti-neutrino flux and the small ionization energy of the
Dresden-II data set, up-scattering via dipole portal interactions can
produce sterile neutrinos with masses up to $\sim 8\,$MeV. In full
generality, one can expect constraints on the effective magnetic
dipole moment coupling to be less severe as the mass of the
up-scattered fermion increases. The kinematic suppression increases,
reaching zero when the sterile neutrino mass hits the kinematic
production threshold limit given by
Eq.~(\ref{eq:kinematic_constraint}). The $1,2,3~\sigma$  (assuming two d.o.f., i.e. $\Delta \chi^2 = 2.3, 6.18, 11.83$ respectively) results of our analysis for this
case are shown Fig.~\ref{fig:dipole-portal}, left (right) graph
obtained with the modified Lindhard (iron-filter) QF. Limits from the
different exclusion regions tend to be a little more uniform in terms
of the up-scattered sterile state mass, in comparison to the NGI
scenarios previously considered in terms of the mediator mass. For the
modified Lindhard QF analysis, values of the order of
 $\mu_{\nu_e}\lesssim 4\times 10^{-10}\,\mu_B$ are excluded for
sterile neutrino masses below 100~keV, at the $1\sigma$
level. Assuming instead the modified iron-filter QF the constraints
are slightly tighter, $\mu_{\nu_e} \lesssim 1 \times 10^{-10}\,\mu_B$ for sterile neutrino
masses below 100~keV, at the $1\sigma$ level.  A comparison of these
values with those obtained using CsI and LAr COHERENT data sets (shown
in the graphs), $\mu_{\nu_e} \lesssim (3-4)\times 10^{-9}\,\mu_B$ at
the $90\%\, $C.L.~\cite{Miranda:2021kre}, demonstrates that the
Dresden-II experimental data improve upon these results (the $90\%$
C.L. upper limits are $(2-8)\times 10^{-10}\,\mu_B$ for
$m_4\lesssim 100\,$keV). They are competitive with the constraints
implied by XENON1T data (indeed more constraining if one focuses only
on the nuclear recoil channel) \cite{Brdar:2020quo}, are stronger than
those derived from CENNS10~\cite{Miranda:2021kre} and comparable (or
even tighter) than those following from TEXONO depending on the QF
model used for the analysis, as can be read directly from the
graphs. If compared with explanations of the XENON1T electron excess
using electron neutrinos \cite{Miranda:2021kre}, one can see that our
results are consistent with that possibility\footnote{Explanations of
  the excess using tau neutrinos are not affected by this result
  either \cite{Shoemaker:2020kji}.}, regardless of the QF choice.
Note that the sterile neutrino dipole portal and NGI results, in
contrast to those found for the weak mixing angle, are to a large
extent rather insensitive to the QF model. Thus, from that point of
view they are more robust.
% ---------------------
% Section: Conclusions
% ---------------------
\section{Conclusions}
\label{sec:conclusions}%
We have studied the implications of the recently released Dresden-II
reactor data on the weak mixing angle and on new physics scenarios
sensitive to the low-energy threshold of the experiment, namely NGI
generated by light vector, scalar and tensor mediators and the sterile
neutrino dipole portal.  In order to check for the dependences on the
QF, we have performed the analyses considering: (i) A modified
Lindhard model, (ii) a QF provided by the collaboration (iron-filter
QF).

The low scintillation energy threshold provides a determination of the
weak mixing angle at a renormalization scale of order $10\,$MeV, a
scale for which up to now no determination was yet available. Our
result shows a rather pronounced dependence on the QF model, with
differences between the best-fit values of about $6\%$. The precision
of the determination of $\sin^2\theta_W$ has
also a strong dependence on that choice, leading to best fit values that are compatible with the SM RGE prediction at $80.7\%$ C.L. and $1\sigma$, respectively. A better understanding
of the germanium QF is thus required to improve upon the determination
of this parameter. However, regardless of these disparities, the
Dresden-II data provides the first hint ever of the value of
$\sin^2\theta_W$ at $\mu \sim 10$ MeV.

Regarding our analysis of NGI with light mediators, also in this case
our findings show that at the $1\sigma$ level results depend on the QF
model. For vector interactions, results derived using the modified
Lindhard QF tend to produce slightly less stringent bounds. In both
cases, though, at large vector mediator masses (above $10\,$MeV or so)
the $1\sigma$ and $2\sigma$ limits produce two nonoverlapping
exclusion regions. At the $3\sigma$ level these regions are gone and
constraints are restricted to a single area, where for vector boson
masses of the order of 100 keV the coupling is constrained to be below
$\sim 10^{-5}$.

The same trend is found for scalar and tensor interactions through
light mediators. Regardless of the QF choice, results lead to
constraints that amount to about $g_{S}\lesssim 1.0\times 10^{-6}$ and $g_{T}\lesssim 1.0\times 10^{-5}$, respectively,
for mediator masses below $\sim 100\,$keV at the $1\sigma$ level. In
all scenarios, the derived constraints turn out to improve upon other
existing bounds from CE$\nu$NS experiments (COHERENT CsI$+$LAr, CONUS
and CONNIE) and even upon predictions made for multi-ton DM detector
measurements.

Finally, concerning the sterile neutrino dipole portal we find that
the Dresden-II results rule out larger regions of parameter space, not
excluded by COHERENT and CONUS and are rather competitive with limits
from XENON1T data. Actually, they are more stringent if one compares
only with XENON1T nuclear recoil data. Compared with those regions
where the sterile neutrino dipole portal can account for the XENON1T
electron excess, the Dresden-II data is not able to test them
yet. However, with more statistics and better understanding of the
germanium QF the situation might improve in the future.

To conclude, the recent evidence for CE$\nu$NS from the Dresden-II
reactor experiment provides unique opportunities to investigate
physics scenarios sensitive to low-energy thresholds, complementary to
other CE$\nu$NS measurements with spallation sources. However, current
results show a dependence on the QF model at low recoil energies thus
calling for a deeper understanding of the germanium QF along with more
data.

\noindent
\textbf{Note added in proof}\\
After completion of the manuscript results from the first science run
of the XENONnT collaboration \cite{XENON:2022mpc} have ruled out the
electron excess previously reported by the XENON1T collaboration
\cite{XENON:2020rca}.
% ------------------
% Acknowledgments
% ------------------
\section*{Acknowledgments}
The authors are grateful to Pilar Coloma, Anirban Majumdar, Sergio
Palomares, J. Collar and I. Katsioulas for useful correspondence.  VDR
acknowledges financial support by the SEJI/2020/016 grant funded by
Generalitat Valenciana, by the Universitat de Val\`encia through the
sub-programme ``ATRACCI\'O DE TALENT 2019'' and by the Spanish grant
PID2020-113775GB-I00 (AEI/10.13039/501100011033). The work of DAS is
supported by ANID grant ``Fondecyt Regular'' N$^\text{o}$ 1221445.

\end{document}